\documentstyle[12pt,cite]{article}
\catcode`\@=11
\renewcommand{\thefootnote}{\fnsymbol{footnote}}

\begin{document}

\begin{titlepage}

{\hfill DFPD/93/TH/72}

\vspace{0.2cm}

{\hfill hep-th/9402081}

\vspace{1.2cm}

{\centerline{{\large \bf NONPERTURBATIVE MODEL OF LIOUVILLE
GRAVITY}\footnote[5]{Partly
supported by the European Community Research
Programme {\it Gauge Theories, applied supersymmetry and quantum
gravity}, contract SC1-CT92-0789}}}

\vspace{1cm}

\centerline{\large{\sc Marco}  {\sc  Matone}\footnote{e-mail:
matone@padova.infn.it, vaxfpd::matone}}

\vspace{0.2cm}

\centerline{\it Department of Physics ``G. Galilei'' -
Istituto Nazionale di Fisica Nucleare}
\centerline{\it University of Padova}
\centerline{\it Via Marzolo, 8 - 35131 Padova, Italy}

\vspace{2.1cm}

\centerline{\large ABSTRACT}

\vspace{0.2cm}

We obtain nonperturbative results in the framework
of continuous Liouville theory.
In particular, we express the specific heat ${\cal Z}$
of pure gravity in terms of an expansion of integrals on
moduli spaces of punctured Riemann spheres. The integrands
are written in terms of the Liouville action. We show that
${\cal Z}$ satisfies the Painlev\'e I.

\end{titlepage}

\newpage

\setcounter{footnote}{0}

\renewcommand{\thefootnote}{\arabic{footnote}}

\noindent
{\bf 1.} In this paper we introduce models of Liouville theory
in the continuum which are based on the Riemann sphere with punctures.
The models include pure gravity.
In particular we will show that
\begin{equation}
{\cal Z}(t)=t^{-12}\sum_{k=4}^\infty t^{5k}\int_{\overline{\cal M}_{0,k}}
\left(i\overline \partial \partial S_{cl}^{(k)}\right)^{k-4}\wedge
\omega^{F_0}-{t^3\over 2}
\label{main1}\end{equation}
is the specific heat of pure gravity, namely ${\cal Z}$ satisfies the
Painlev\'e I
\begin{equation}
{\cal Z}^2(t)-{1\over 3}{\cal Z}''(t)=t.
\label{PI}\end{equation}
$S_{cl}^{(k)}$ in (\ref{main1}) denotes the {\it classical}
Liouville action on the $k$-punctured Riemann sphere.
The class $[\omega^{F_0}]$ is the Poincar\'e dual of a divisor on
the compactified moduli space $\overline{\cal
M}_{0,k}$ which is given in terms of the $(2k-8)$-cycles defining
the Deligne-Knudsen-Mumford boundary of $\overline{\cal M}_{0,k}$.
The basic tools to obtain (\ref{main1}) are classical Liouville
theory and intersection theory.

This result reproduces in the continuum the well-known result
obtained in the matrix model approach to pure gravity \cite{mm}.
For reviews on matrix models and 2D gravity see \cite{mm2}.

\vspace{0.3cm}

\noindent
{\bf 2.}
The problems arising in the continuum formulation
of Liouville gravity \cite{aa,a,b} are essentially:
\begin{itemize}
\item[{\bf a.}]{To evaluate Liouville correlators
on Riemann surfaces of genus $h\ge 2$;}
\item[{\bf b.}]{To perform the integration on moduli spaces;}
\item[{\bf c.}]{To recover nonperturbative results from the
topological expansion.}
\end{itemize}
 Results from matrix models and topological
gravity show that these aspects are strictly related with the structure
of ${\cal M}_h\equiv {\cal M}_{h,0}$
(we denote by ${\cal M}_{h,n}$ the moduli spaces of Riemann surfaces of genus
$h$ and $n$ punctures).
In particular,
it turns out that the Liouville action is the K\"ahler potential for the
natural (Weil-Petersson) metric on the moduli space. Also CFT is
strictly related with the geometry of moduli space.
For example, the Mumford isomorphism
$$
\lambda_n\cong \lambda_1^{c_n},\qquad c_n=6n^2-6n+1,
$$
where $\lambda_n=\det \,{\rm ind}\,\overline \partial_n$
are the determinant line bundles, connects  geometrical properties
of ${\cal M}_h$ with the central charge  $d=-2c_n$ of a weight $n$, $b$-$c$
system (notice that $d\le 1$). Actually, the bosonization of $b$-$c$
systems can be used to reproduce the Coulomb gas formulation of
$d\le 1$ conformal matter.
For $d>1$ it is not possible to represent conformal matter
by a $b$-$c$ system. In this case
one can consider the  $\beta$-$\gamma$
system of weight $n$ whose central charge is  $2c_n$. However,
the representation of the $\beta$-$\gamma$ system in terms of free
fields is a long-standing problem which seems related to the
$d=1$ barrier.
These aspects indicate that there is a connection between
the barrier and the Mumford isomorphism. This
is related to a similar structure
considered in \cite{mmb} in the framework of the geometrical formulation
of 2D gravity \cite{LAT,mmb} where representing elliptic and parabolic
Liouville operators by means of a scalar field constrains
the conformal matter to be in the sector $d\le 1$.

The natural framework to investigate the aspects considered above is
the theory of uniformization of Riemann surfaces
where Liouville theory plays a crucial role. Actually, in \cite{grava}
it has been shown that the Liouville action appears in the
correlators (intersection numbers) of topological gravity \cite{1}.
The relationships between Liouville theory, matrix models and topological
gravity suggest that it is possible to extend the above
Liouville-topological gravity relationship
by recovering the nonperturbative results of matrix models by continuum
Liouville theory. In our model we will reduce all aspects
concerning higher genus contributions to punctured spheres. The
reduction to punctured sphere has been considered also
by V.G. Knizhnik who expressed the sum of the genus expansion as a CFT on an
arbitrary $N$-sheet covering of the Riemann sphere with branch points.
For each branch point he associated a vertex
operator and proposed to express the infinite sum on all genus ($h\ge
2$) as the limit for $N\to\infty$ of a `nonperturbative' partition
function \cite{Knizhniknonperturbative}.

A natural way to get punctured spheres is
by pinching all handles of a compact Riemann surface.
Degenerate
(singular) surfaces
belong to the boundary of moduli spaces.
These singularities
play a fundamental role
in the evaluation of relevant integrals (intersection theory).
 The fact
that the {\it classical} Liouville action is the K\"ahler potential for the
Weil-Petersson metric and the structure of the boundary of
moduli space suggest to
consider integrals on $\overline{\cal M}_h$ in the framework
of the Duistermaat-Heckman integration formula \cite{DuisterHeckm}.
The final result should be a sum of integrals $Z_n^{F}$ on the moduli space of
punctured Riemann spheres $\overline{\cal M}_{0,n}=
\left(\widehat{\bf C}\backslash\Delta_n\right)/Symm(n)\times PSL(2,{\bf
C})$
with the integrands involving the Liouville action.
These remarks indicate that a theory \`a la Friedan-Shenker
\cite{FriedanShenker} can be concretely
formulated to recover nonperturbative results
in the continuum formulation.
In this paper, we do not consider points {\bf a}-{\bf c} separately,
rather we state the final solution finding
the explicit form of the integrals $Z_n^{F}$ on
$\overline{\cal M}_{0,n}$.

 The reduction to punctured sphere is
particularly evident in topological field theory
coupled to 2D gravity where higher genus contributions
to the free energy $\langle 1\rangle_h$
can be written in terms of the sphere amplitudes of the
puncture operator $P$ \cite{1,hsl}.
The physical
observables of the theory are the primary fields
${\cal O}_\alpha$ ($\alpha=0,1,\ldots,N-1$, ${\cal O}_0$ is the identity
operator) and their gravitational descendents
$\sigma_n\left({\cal O}_\alpha\right)$, $n=1,2,\ldots$. In the coupled system
${\cal O}_0$ becomes non-trivial and it is identified with $P$. Denoting by
${\cal L}_0$ the minimal Lagrangian, the more general one is
${\cal L}={\cal L}_0+\sum_{n,\alpha}t_{n,\alpha}\sigma_n\left({\cal O}_\alpha
\right),\sigma_0\left({\cal O}_\alpha\right)\equiv {\cal
O}_\alpha,$
where $t_{n,\alpha}$ are coupling constants. With this definition one can
compute correlation functions with an insertion of $\sigma_k$
just by differentiating $\langle 1\rangle_h$ with respect to $t_k$.
Thus in general
 $$
\big <\sigma_{d_1}\left({\cal O}_{\alpha_1}\right)\cdot\cdot\cdot\sigma_{d_n}
\left({\cal O}_{\alpha_n}\right)\big >_h={\partial\over \partial
t_{d_1,\alpha_1}}\cdot\cdot\cdot{\partial\over \partial
t_{d_n,\alpha_n}}\big <1\big >_h.
$$
Therefore $\big <1\big >_h$ is the crucial quantity
to compute.
By means of KdV recursion relations
$$
\big <\sigma_1(P)P\big >_h=2\big <P^4\big >_{h-1}+
{1\over 2}\sum_{h'=0}^h\big <P^2\big >_{h'}\big <P^2\big >_{h-h'},
$$
it is possible  \cite{hsl}
to express $\big <1\big >_h$ as a sum of terms of the
 form $\displaystyle \big <P^{n_1}\big >_0\cdot\cdot\cdot
\big <P^{n_j}\big >_0/\big <P^3\big >_0^{h+j-1}$ for $1\le j \le 3h-3$
with  the constraint $\sum_{k=1}^j n_k=3(j+h-1)$.

The reduction to the punctured sphere arises also
in the evaluation of ${\rm Vol}_{WP}({\cal M}_{h,n})$.
Indeed, at least in some cases, there is  a relationship between
$\overline{\cal M}_{h,n}$, $\overline{\cal M}_{0,n+3h}$ and their
volumes\footnote{The space
${\cal M}_{h,n}$ is not affine for $h>2$. Conversely
the space ${\cal M}_{0,k}$ is finitely covered by the affine space
$V^{(k)}$ defined in (\ref{star}). Thus for $h>2$
the are not geometrical isomorphisms between
$\overline{\cal M}_{h,n}$ and $\overline{\cal M}_{0,n+3h}$.
However, in principle, nothing exclude the possibility to express
${\rm Vol}_{WP}\left({\cal M}_{h,n}\right)$ in terms of
${\rm Vol}_{WP}\left({\cal M}_{0,n+3h}\right)$.}.
The first example is the geometric isomorphism \cite{wolpertis}
$\overline{\cal M}_{1,1}\cong \overline{\cal M}_{0,4}$,
and
\begin{equation}
{\rm Vol}_{WP}\left({\cal M}_{1,1}\right)
=2{\rm Vol}_{WP}\left({\cal M}_{0,4}
\right).
\label{wolpertvolume}\end{equation}
To understand this result it is sufficient to
recall that the $\wp$-function
enters in the expression of the uniformizing connection
of the once punctured torus $\Sigma_{1,1}$
(note that $\wp$ is a solution of the KdV equation)
$$
T_{\Sigma_{1,1}}={1\over 2}\left(\wp(\tau,z)+c(\tau)\right),
$$
where $c(\tau)$ is the accessory parameter for $\Sigma_{1,1}$.
Eq.(\ref{wolpertvolume}) follows  from the fact that
$T_{\Sigma_{1,1}}$ is strictly related to the uniformizing connection
$T_{\Sigma_{0,4}}$ of the Riemann sphere with four punctures
since $\wp$ maps $\Sigma_{1,1}$ two-to-one onto
the four punctured Riemann sphere.
Let us notice that another isomorphism  is \cite{igusa}
$\overline{\cal M}_{2,0}\cong \overline{\cal M}_{0,6}$.

There is another way to understand why punctured spheres play a crucial
role in 2D gravity. The point is to notice that equal size
triangulated Riemann surfaces considered in matrix models
can be realized in terms of
thrice punctured spheres \cite{LevinMorozov}.
This aspect is related to arithmetic surfaces theory
\cite{LevinMorozov,SMIT}. In this
context one should investigate whether this kind of surfaces have some
suitable symmetry to define
antiholomorphic involution.
 This question is
important in order to investigate Osterwalder-Schrader positivity.
This is connected with the problem of defining the adjoint in higher
genus.
On the sphere it can be done thanks to the
natural antinvolution $z\to \bar z^{-1}$. In higher genus
this problem has been solved only
on a Schottky double where there is a natural antinvolution
\cite{jkl}.
Recently Harvey and Gonz\'alez Di\'ez \cite{GabinoHarvey}
have considered loci of curves
which are prime Galois covering of the sphere. In particular they
considered the important case of Riemann surfaces admitting non-trivial
automorphisms and showed that there is a birational isomorphism between
a subset of the moduli space ${\cal M}_h$ and $V^{(n)}$ (defined in
(\ref{star})).

\vspace{0.3cm}

{\bf 3.} The relation between Liouville and uniformization theory of Riemann
surfaces arises in considering
the Liouville equation
\begin{equation}
\partial_{\bar
z}\partial_z\varphi_{cl}={e^{\varphi_{cl}}\over 2},
\label{le}\end{equation}
 which is uniquely satisfied
by the Poincar\'e metric (i.e. the metric with Gaussian curvature
$-1$). This metric can be written in terms of the inverse of the
uniformizing map $J_H$, that is
$e^{\varphi_{cl}}={|{J_H^{-1}}'|^2/ \left({\rm Im}\,
J_H^{-1}\right)^2}$, $J_H:H\to \Sigma\cong H/\Gamma$
where $H$ is the upper half-plane and $\Gamma$ a Fuchsian group.
Let us introduce the $n$-punctured sphere
$\Sigma=\widehat
{\bf C}\backslash\{z_1,\ldots,z_n\},
\widehat {\bf C}\equiv {\bf C}\cup\{\infty\}$.
Its moduli space is the
space of classes of isomorphic $\Sigma$'s, that is
\begin{equation}
{\cal M}_{0,n}=
\{(z_1,\ldots,z_{n})\in
\widehat{\bf C}^{n}|z_j\ne z_k\; {\rm for}\; j\ne k\}/Symm(n)\times
PSL(2,{\bf C}), \label{modulisp}\end{equation}
where ${Symm}(n)$ acts by permuting
$\{z_1,\ldots,z_n\}$ whereas $PSL(2,{\bf C})$ acts by linear fractional
transformations. By $PSL(2,\bf C)$ we can recover the `standard
normalization':  $z_{n-2}=0$, $z_{n-1}=1$ and $z_{n}=\infty $.
Furthermore, without loss of generality, we assume
that $w_{n-2}=0$, $w_{n-1}=1$ and
$w_n=\infty$.
  For the classical Liouville tensor we have
$$
T^F(z)
=\sum_{k=1}^{n-1}\left({1\over 2(z-z_k)^2}+
{c_k\over z-z_k}\right),\qquad
\lim_{z\to \infty}T^F(z)={1\over 2z^2}+{c_n\over z^3}+
{\cal O}\left({1\over |z|^4}\right),
$$
with the following constraints on the
{\it accessory parameters}
$$
\sum_{k=1}^{n-1}c_k=0,  \qquad \sum_{k=1}^{n-1}c_kz_k=1-n/2,
\qquad \sum_{k=1}^{n-1}z_k(1+c_kz_k)=c_n.
$$
The $c_k$'s are functions on
\begin{equation}
 V^{(n)}=\{(z_1,\ldots,z_{n-3})\in
{\bf C}^{n-3}|z_j\ne 0,1; z_j\ne z_k,\; {\rm for}\; j\ne k\}.
\label{star}\end{equation}
Note that
\begin{equation}
{\cal M}_{0,n}\cong V^{(n)}/{Symm}(n),
\label{mdls}\end{equation}
where the action of $Symm(n)$ on $V^{(n)}$ is defined by comparing
(\ref{modulisp}) with (\ref{mdls}).

Let us now consider the compactification divisor (in the sense of
Deligne-Knudsen-Mumford)
$D=\overline V^{(n)}\backslash V^{(n)}$.
This divisor decomposes
in the sum of divisors $D_1$,\ldots,$D_{[n/2]-1}$ which are subvarieties
of real dimension  $2n-8$. The locus $D_k$
consists of surfaces that split, on  removal of the node, into two
Riemann spheres with $k+2$ and $n-k$ punctures.
In particular $D_k$ consists of $C(k)$ copies of the space
$\overline V^{(k+2)}\times \overline V^{(n-k)}$ where
$C(k)=\left(^{\;\; n}_{k+1}\right)$ for $k=1,\ldots,{(n-1)\over 2}-1$,
with the exception that for $n$ even
$C(n/2-1)={1\over 2}\left(^{\;\; n}_{n/2}\right)$.
It turns out that the image of the $D_k$'s, provide
a basis in $H_{2n-8}(\overline{\cal M}_{0,n},{\bf R})$.

In the case of
the punctured Riemann sphere  eq.(\ref{le}) follows from the Liouville action
$$
S^{(n)}=\lim_{r\to 0}\left[\int_{\Sigma_r}
\left(\partial_z\varphi\partial_{\bar z}{\varphi}+e^{\varphi}\right)+
2\pi (n {\log} r+2(n-2){\log}|{\log}r|)\right],
$$
where $\Sigma_r=\Sigma\backslash\left(\bigcup_{i=1}^{n-1}
\{z||z-z_i|<r\}\cup\{z||z|>r^{-1}\}\right)$.
This action, evaluated on the classical solution,
is the K\"ahler potential for the Weil-Petersson two-form
on $V^{(n)}$ \cite{0}
\begin{equation}\omega_{WP}^{(n)}= {i\over
2}{\overline\partial}{\partial}S^{(n)}_{cl}=-i\pi\sum_{j,k=1}^{n-3}
 {\partial c_k\over \partial {\bar z_j}}d\bar z_j\wedge d z_k.
\label{36}\end{equation}

Let us consider the volume of moduli space of punctured Riemann
spheres
$$
{\rm Vol}_{WP}\left({\cal M}_{0,n}\right)={1\over (n-3)!}
\int_{\overline{\cal M}_{0,n}}{\omega_{WP}^{(n)}}^{n-3}=
{1\over (n-3)!}
\left[\omega_{WP}^{(n)}\right]^{n-3}\cap
\left[\overline{\cal M}_{0,n}\right].
$$
Recently it has been shown that \cite{01}
$$
{\rm Vol}_{WP}\left({\cal M}_{0,n}\right)={1\over n!}
{\rm Vol}_{WP}\left(V^{(n)}\right)={\pi^{2(n-3)}
 V_n\over n!(n-3)!},\qquad n\ge 4,
$$
where $V_n=\pi^{2(3-n)}\left[\omega_{WP}^{(n)}\right]^{n-3}\cap
\left[\overline{V}^n\right]$
satisfies the recursion relations
\begin{equation}   V_3=1,\qquad
V_n={1\over 2}\sum_{k=1}^{n-3}{k(n-k-2)
\over n-1}\left(^{\;\;n}_{k+1}\right)\left(^{n-4}_{k-1}\right)
 V_{k+2}V_{n-k},\qquad
 n \ge 4.\label{51}\end{equation}
Remarkably the basic tools in the computation of the volumes
are classical Liouville theory and intersection theory.

\vspace{0.3cm}

\noindent
{\bf 4.} We now consider the  differential equation associated with (\ref{51}).
First of all we define
\begin{equation}
 a_k= {V_k\over (k-1)((k-3)!)^2},\qquad k\ge 3,
\label{rnm2}\end{equation}
so that (\ref{51}) becomes
\begin{equation}
a_3=1/2,\qquad
a_n={1\over 2}{n(n-2)\over (n-1)(n-3)}
\sum_{k=1}^{n-3}a_{k+2}a_{n-k},\qquad n\ge 4.\label{51al}\end{equation}
Eq.(\ref{51al}) is equivalent to the differential equation
\begin{equation}
g''={{g'}^2t-gg'+g't\over t(t-g)},
\label{51a}\end{equation}
where $g(t)=\sum_{k=3}^\infty a_k t^{k-1}$.
Notice that by (\ref{36})
\begin{equation}
g(t)=\sum_{k=3}^\infty {k(k-2)\over (k-3)!}t^{k-1}
\int_{\overline{\cal M}_{0,k}}
\left({i\overline\partial\partial
S_{cl}^{(k)}\over 2\pi^2}\right)^{k-3},
\label{prtbtv}\end{equation}
where ``${\int_{\overline{\cal M}_{0,3}}1}$''$\equiv {1\over 6}$.
Function $g(t)$ resembles a topological expansion of string theory.
Furthermore the structure of eq.(\ref{51a}) resembles the Painlev\'e
I. These remarks indicate that it is possible to recover the specific
heat of pure gravity in the continuum.
Actually, we will recover the Painlev\'e I by classical Liouville theory.
In particular we will get the recursion relations for the
Painlev\'e I by performing a suitable modification of the
Weil-Petersson volume form ${\omega_{WP}^{(n)}}^{n-3}$.
Remarkably, as we will show,
it is possible to perform the substitution
$$
{\omega_{WP}^{(n)}}^{n-3}\longrightarrow {\omega_{WP}^{(n)}}^{n-4}
\wedge \omega^F,
$$
in (\ref{prtbtv}) without changing the general structure of
(\ref{51al}); that is we will get recursion relations of the following
structure
\begin{equation}
A_n=C(n)\sum_{k=1}^{n-3}A_{k+2}A_{n-k},\qquad n\ge 4.
\label{main3}\end{equation}

The first problem is to find a suitable expansion for the Painlev\'e
I field such that the structure of the
associated recursion relation be the same of (\ref{main3}).
Remarkably this expansion exists, namely
\begin{equation}
f(t)=t^{-12}\sum_{k=3}^\infty d_k
t^{5k}.\label{painlfield}\end{equation}
It is interesting that in searching the expansion reproducing the
general structure of (\ref{51al}), which is a result obtained from continuous
Liouville theory, one obtains
an expansion involving only {\it positive} powers
of $t$. With this expansion the Painlev\'e I
\begin{equation}
f^2(t)-{1\over 3}f''(t)=t,\label{painleve}\end{equation}
is equivalent to the recursion relations\footnote{Notice
 that $(-1)^kd_k$ is positive.}
\begin{equation}
d_n={3\over (12-5n)(13-5n)}\sum_{k=1}^{n-3} d_{k+2}d_{n-k},
\qquad d_3=-1/2,\label{rr12}\end{equation}
which has the same structure of
(\ref{51al}). We now investigate on the possible volume forms
reproducing (\ref{rr12}). To understand which kind of modification to
${\omega_{WP}^{(n)}}^{n-3}$ can be performed without changing the basic
structure of (\ref{51al}) we recall basic steps in \cite{01}
to obtain (\ref{51}).

Let $D_{WP}$ be the $(2n-8)$-cycle dual
to the Weil-Petersson class
$\left[\omega_{WP}^{(n)}\right]$.
To compute the volumes it is useful to expand
$D_{WP}$ in terms of the divisors $D_k$ in the boundary of moduli space.
It turns out that \cite{01}
\begin{equation}
D_{WP}={\pi^2\over n-1}\sum_{k=1}^{[n/2]-1}k(n-k-2)D_k.
\label{wpdvsr}\end{equation}
Let us set
$$
\widetilde V_n=\pi^{2(n-3)}V_n=
\left[\omega_{WP}^{(n)}\right]^{n-3}\cap\left[\overline
V^{(n)}\right]=
\left[\omega_{WP}^{(n)}\right]^{n-4}\cap
\left(\left[\omega_{WP}^{(n)}\right]\cap\left[\overline
V^{(n)}\right]\right).
$$
On the other hand $\left[\omega_{WP}^{(n)}\right]\cap\left[\overline
V^{(n)}\right]=D_{WP}\cdot \overline
V^{(n)}=D_{WP}$, so that by (\ref{wpdvsr})
$$
\widetilde
V_n=\left[\omega_{WP}^{(n)}\right]^{n-4}\cap\left[D_{WP}\right]=
{\pi^2\over n-1}\sum_{k=1}^{[n/2]-1}k(n-k-2)
\left[\omega_{WP}^{(n)}\right]^{n-4}\cap\left[D_k\right].
$$
Since $D_k$ consists of $C(k)$ copies of the space
$\overline V^{(k+2)}\times \overline V^{(n-k)}$, we have
$$
\widetilde V_n={\pi^2\over n-1}\sum_{k=1}^{[n/2]-1}
k(n-k-2)C(k)
\left[\omega_{WP}^{(n)}\right]^{n-4}\cap
\left[\overline V^{(k+2)}\times \overline V^{(n-k)}\right].
$$
Finally, since \cite{01}
\begin{equation}
\left[\omega_{WP}^{(n)}\right]^{n-4}\cap
\left[\overline V^{(k+2)}\times \overline V^{(n-k)}\right]=
\left[\omega_{WP}^{(k+2)}+\omega_{WP}^{(n-k)}\right]^{n-4}\cap
\left[\overline V^{(k+2)}\times \overline V^{(n-k)}\right],
\label{ab2}\end{equation}
it follows that
$$
\widetilde V_{3}=1,\qquad
\widetilde V_{n}={\pi^2\over n-1}\sum_{k=1}^{[n/2]-1}
k(n-k-2)C(k)\left(^{n-4}_{k-1}\right)
\widetilde V_{k+2} \widetilde V_{n-k},\qquad n\ge 4,
$$
which coincides with (\ref{51}).

We now introduce the divisor
\begin{equation}
D^F={\pi^2\over n-1}\sum_{k=1}^{[n/2]-1}k(n-k-2)F(n,k)D_k,
\label{wpdvsrF}\end{equation}
where $F(n,k)$ is a function to be determined. Let
$[\omega^F]$ be the Poincar\'e dual to $D^F$ and define
\begin{equation}
Z_n^F=
\int_{\overline{\cal M}_{0,n}}{\omega_{WP}^{(n)}}^{n-4}
\wedge \omega^F=
\int_{\overline{\cal M}_{0,n}}
\left({i\overline\partial\partial
S_{cl}^{(n)}\over 2}\right)^{n-4}
\wedge \omega^F,\qquad  n\ge 4.
\label{48d}\end{equation}
An important aspect of (\ref{48d}) is that we can use the recursion
relations (\ref{51al}) to obtain nonperturbative results. This
possibility is based on the obvious, but important fact, that
$\left[\omega_{WP}^{(n)}\right]^{n-3}\cap\left[\overline V^{(n)}\right]=
\left[\omega_{WP}^{(n)}\right]^{n-4}\cap\left[D_{WP}\right]$ implying
that the general structure of (\ref{51al}) (the same of (\ref{rr12})) is
unchanged under the substitution ${\omega_{WP}^{(n)}}^{n-3}
\longrightarrow {\omega_{WP}^{(n)}}^{n-4}\wedge \omega^F$. To see this
note that
\begin{equation}
Z^F_n={1\over n!}\left[\omega_{WP}^{(n)}\right]^{n-4}\cap
\left[D^F\right]
={\pi^2\over (n-1)n!}\sum_{k=1}^{[n/2]-1}F(n,k)k(n-k-2)
\left[\omega_{WP}^{(n)}\right]^{n-4}\cap\left[D_k\right],
\label{aa}\end{equation}
On the other hand by (\ref{ab2})
\begin{equation}
\sum_{k=1}^{[n/2]-1}F(n,k)k(n-k-2)
\left[\omega_{WP}^{(n)}\right]^{n-4}\cap\left[D_k\right]=
{1\over 2}\sum_{k=1}^{n-3}F(n,k){k(n-k-2)\over n-1}
\left(^{\;\; n}_{k+1}\right)\left(^{n-4}_{k-1}\right)V_{k+2}V_{n-k},
\label{adsf}\end{equation}
and by (\ref{rnm2})
\begin{equation}
Z_n^F={\pi^2\over 2}{(n-4)!\over n-1}\sum_{k=1}^{n-3}F(n,k)
a_{k+2}a_{n-k},\qquad n\ge 4.
\label{basg}\end{equation}

Let us define the `Liouville $F$-models'
\begin{equation}
{\cal Z}^{F,\alpha}(x)=x^{-\alpha}\sum_{k=3}^{\infty}x^kZ_k^F,
\label{Fmodels}\end{equation}
where $x$ is the coupling constant. These models are classified by $\alpha$,
$F(n,k)$ and $Z_3^F$.
 We now show that ${\cal Z}^{F,\alpha}(x)$ includes pure gravity.
In fact, putting
\begin{equation}
{\cal Z}(t)={\cal Z}^{F_0,\alpha_0}(t^5), \qquad Z_3^{F_0}=-1/2,
\quad \alpha_0=12/5,
\label{PIa}\end{equation}
where
\begin{equation}
F_0(n,k)=
{6\over \pi^2}{(n-1)\over (12-5n)(13-5n)(n-4)!}
{Z_{k+2}^{F_0}Z_{n-k}^{F_0}\over a_{k+2}a_{n-k}},
\label{setnumber}\end{equation}
we have, by (\ref{basg}) and (\ref{setnumber}),
\begin{equation}
Z_3^{F_0}=-1/2,\qquad
Z_n^{F_0}={3\over
(12-5n)(13-5n)}\sum_{k=1}^{n-3}Z_{k+2}^{F_0}Z_{n-k}^{F_0},\qquad
n\ge 4,
\label{star2}\end{equation}
so that by (\ref{painlfield})-(\ref{rr12})
\begin{equation}
{\cal Z}(t)=t^{-12}\sum_{k=4}^\infty t^{5k}
\int_{\overline{\cal M}_{0,k}}
\left({i\overline\partial\partial
S_{cl}^{(k)}\over 2}\right)^{k-4}
\wedge \omega^F-{t^3\over 2},
\label{ppII}\end{equation}
satisfies the Painlev\'e I
\begin{equation}
{\cal Z}^2(t)-{1\over 3}{\cal Z}''(t)=t.
\label{PPPP}\end{equation}

\vspace{0.3cm}

{\bf 5.} In conclusion we have introduced a class of Liouville models
by defining a suitable $D^F$ divisor. These Liouville $F$-models (LFM)
include pure gravity. In this context we recall that the Liouville
action arises also in the correlators of topological gravity
\cite{grava}.

Punctures correspond to real points on the boundary of the upper
half-plane. Correspondingly one can define hyperelliptic Riemann surfaces.
In the case of infinite genus one gets the McKean-Trubowitz
\cite {MT} model which is related to matrix model.
This suggests a nonperturbative formulation on $H$ with the image
of punctures related to the eigenvalues of the Hermitian matrix models.
In the discrete version of this
approach one should be able to connect this formulation
with the ideas at the basis of \cite{BX}.

\vspace{0.5cm}

{\bf Acknoweledgements}
I would like to thank G. Bonelli for stimulating discussions.


\begin{thebibliography}{99}

\bibitem{mm} E. Br\'ezin and V. Kazakov, Phys. Lett. {\bf 236B} (1990) 144.
M. Douglas and S. Shenker, Nucl. Phys. {\bf B335} (1990) 635.
D. Gross and
A. Migdal, Phys. Rev. Lett. {\bf 64} (1990) 127.
\bibitem{mm2} L. Alvarez-Gaum\'e, {\it Random Surfaces, Statistical
Mechanics And String Theory}, Lausanne lectures, winter 1990.
P. Ginsparg, {\it Matrix Models Of 2D Gravity}, Trieste Lectures,
LA-UR-91-4101, hepth/9112013. E. Martinec, {\it An Introduction To 2D
Gravity And Solvable String Models}, Trieste Lectures, RU-91-51,
hepth/9112019. A. Morozov, {\it Integrability And Matrix Models},
ITEP-M2/93, hepth/9303139. P. Di Francesco, P. Ginsparg and J.
Zinn-Justin, {\it 2D Gravity And Random Matrices}, LA-UR-93-1722,
SPhT/93-061, hepth/9306153.
\bibitem{aa} A.M. Polyakov, Phys. Lett. {\bf 103B} (1981) 207.
\bibitem{a} F. David, Mod. Phys. Lett {\bf A3} (1988) 509.
J. Distler and H. Kawai, Nucl. Phys. {\bf B321} (1989) 509.
\bibitem{b} J.-L. Gervais, Comm. Math. Phys. {\bf 130} (1990) 257;
{\bf 138} (1991) 301.
\bibitem{mmb} M. Matone, {\it Quantum Riemann Surfaces, 2D Gravity And
The Geometrical Origin Of Minimal Models}, preprint DFPD/93/TH/62,
hepth/9309096.
\bibitem{LAT} L.A. Takhtajan, {\it Liouville Theory: Quantum
Geometry Of Riemann Surfaces}, hepth/9308125.
\bibitem{grava} M. Matone, {\it Uniformization Theory and 2D Gravity.
I. Liouville Action And Intersection Numbers},
IC-MATH/8-92, DFPD-TH/92/41, hepth/9306150.
\bibitem{1} E. Witten, Nucl. Phys. {\bf B340} (1990) 281;
 Surv. Diff. Geom. {\bf 1} (1991) 243.
R. Dijkgraaf and E. Witten, Nucl. Phys. {\bf B342} (1990) 486.
\bibitem{Knizhniknonperturbative} V.G. Knizhnik, Sov. Phys. Usp.
{\bf 32} (11) (1989) 945.
\bibitem{DuisterHeckm} J.J. Duistermaat and G.J. Heckman, Invent. Math.
{\bf 69} (1982) 259; {\bf 72} (1983) 153. N. Berline and M. Vergne,
Duke Math. J. {\bf 50} (1983) 539. R.F. Picken J. Math. Phys. {\bf 31}
(1990) 616.
\bibitem{FriedanShenker} D. Friedan and S. Shenker, Nucl. Phys.
{\bf B281} (1987) 509.
\bibitem{hsl} J.H. Horne and S.P. Martin,
Phys. Lett. {\bf 258B} (1991) 322. A. Sen,
Int. J. Mod. Phys. {\bf A7} (1992) 2559. K. Li,
Nucl. Phys. {\bf B354} (1991) 725.
\bibitem{wolpertis}
 S.A. Wolpert, Ann. of Math. {\bf 118} (1983) 491.
\bibitem{igusa} J. Igusa, Ann. of Math. {\bf 72} (1960) 612.
\bibitem{LevinMorozov} A. Levin and A. Morozov, Phys. Lett. {\bf 243B}
(1990) 207.
\bibitem{SMIT} D.-J. Smit, Comm. Math. Phys. 143 (1992) 253.
\bibitem{jkl} A. Jaffe, S. Klimek and A. Lesniewski,
Comm. Math. Phys. {\bf 126} (1989) 421.
\bibitem{GabinoHarvey} G. Gonz\'alez-Di\'ez and W.J. Harvey,
{\it Moduli Of Riemann Surfaces With Symmetry}, in Discrete Groups and
Geometry, ed. W.J. Harvey and C. Maclachlan, Cambridge 1992.
G. Gonz\'alez-Di\'ez Proc. London Math. Soc. 62 (1991) 469;
{\it On Prime Galois Covering Of The Riemann Sphere}, preprint.
\bibitem{0} P.G. Zograf and L.A. Takhtajan,
Math. USSR Sbornik, {\bf 60} (1988) 143; {\bf 60} (1988) 297.\\ L.A. Takhtajan,
Proc. Symp.  Pure Math. (AMS)
{\bf 49}, part 1 (1989) 581.
\bibitem{01} P.G. Zograf, {\it The Weil-Petersson
Volume Of The Moduli Space Of Punctured
Spheres}, to appear in Cont. Math..
\bibitem{MT} H.P. McKean and E. Trubowitz, Comm. Pure Apll. Math.
{\bf 29} (1976) 143.
\bibitem{BX} L. Bonora and C.S. Xiong, {\it Correlation Functions Of
Two-Matrix Models}, SISSA 172/93/EP, BONN-HE-45/93, hepth/9311089.
\end{thebibliography}
\end{document}